# On unorthodox qubits, with an application to the closed timelike curve problem

By Samuel Kuypers[1]

**Abstract**

In orthodox quantum theory the observables of spacelike separated quantum systems commute. I shall call this the commutation constraint. It severely limits quantum theory's explanatory power. For instance, the constraint cannot be met in the presence of closed timelike curves, leaving us with no choice but to rule them out by fiat. It also conflicts with Bekenstein's bound. Here I investigate a modified quantum theory, unorthodox quantum theory, which is different from the conventional theory only in its omission of this commutation constraint. In particular, I describe a system of unorthodox qubits and demonstrate how they can be used to model systems on closed timelike curves and how they allow for a solution of the grandfather paradox.

## 1 The commutation constraint

In what I shall call *orthodox quantum theory*, systems at spacelike separations always have commuting descriptors. So, in the Heisenberg picture, if $\hat{A}(t)$ and $\hat{B}(t)$ are descriptors of systems $\mathfrak{A}$ and $\mathfrak{B}$ respectively at time $t$, then

$$\left[\hat{A}(t), \hat{B}(t)\right] = 0. \tag{1.1}$$

This *commutation constraint* is imposed both in non-relativistic quantum theory and in quantum field theory.[2] Its main virtue is that it imposes Einsteinian locality (Deutsch & Hayden (2000), and Bédard (2021)), which is the requirement that 'the real factual situation of [system $\mathfrak{A}$] is independent of what is done with [system $\mathfrak{B}$], which is spatially separated from [$\mathfrak{A}$]' (Einstein (1949), p. 85). That is to say, it ensures that the descriptors of physical systems only affect one another locally.[3]

[1] The Clarendon Laboratory, University of Oxford, Oxford OX1 3PU, UK

[2] The assumption that spacelike separated *fermionic* systems should anti-commute suffers from the same problems as the commutation constraint. Therefore I shall confine attention to the commutation constraint while bearing in mind that any other fixed algebraic-relation between spacelike separated systems is similarly problematic.

[3] As has been demonstrated by Deutsch & Hayden (*op. cit.*), quantum theory satisfies Einstein locality, while also being Bell non-local, meaning that quantum theory does not admit a local hidden-variable model in terms of real-valued stochastic observables.

However, the commutation constraint is merely a sufficient, but not a necessary, requirement for Einsteinian locality: there exist Einstein-local theories that violate (1.1) (cf. Marletto *et al.* (2021)). Moreover, the commutation constraint is untenable in various situations, which severely limits orthodox quantum theory's explanatory power. I shall discuss two such situations: spacetimes with closed timelike curves; and quantum field theory.

### 1.1 Closed timelike curves

For a quantum system to abide by the commutation constraint, spacetime must admit foliation by spacelike hypersurfaces (see for instance Hawking (1995) and Hartle (1994)). Yet, such foliations do not exist when spacetime contains closed timelike curves – for instance, the Gödel (1949) metric, the Kerr metric (Hawking & Ellis (1973), ch. 5), and spacetimes with traversable wormholes (Morris *et al.* (1988), and Lockwood (2007), ch. 6). This is a problem even if those spacetimes are not realised in nature because general relativity does not rule out the existence of closed timelike curves while orthodox quantum theory does. Hence, because of the commutation constraint there is a potential conflict between those two theories, which would have to be resolved by fiat.

To demonstrate the conflict, I shall describe perhaps the simplest quantum system, a qubit $\mathfrak{Q}_1$, existing on a closed timelike curve. Following Deutsch & Hayden (*op. cit.*) and Gottesman (1999), $\mathfrak{Q}_1$ at a general time $t$ is represented by a triple of q-number descriptors

$$\hat{\boldsymbol{q}}_1(t) = \left(\hat{q}_{1x}(t),\ \hat{q}_{1y}(t),\ \hat{q}_{1z}(t)\right), \tag{1.2}$$

which at all times $t$ satisfy the Pauli algebra

$$\left.\begin{aligned} \hat{q}_{1i}(t)\hat{q}_{1j}(t) &= \delta_{ij}\hat{1} + i\epsilon_{ij}{}^{k}\hat{q}_{1k}(t) \\ \hat{q}_{1i}(t)^{\dagger} &= \hat{q}_{1i}(t) \end{aligned}\right\} \qquad (i,j,k \in \{x,y,z\}). \tag{1.3}$$

Here † denotes Hermitian conjugation. I adopt Einstein's summation convention: when an index appears twice in a product, once as a superscript and once as a subscript, the expression is implicitly summed over all its possible values.

I will treat the closed timelike curve as two spacelike cuts in spacetime, whose edges are identified, and I will assume that outside of the chronology-violating region there exists an unambiguous future and past (figure 1.a). I will abstract away the remaining spacetime geometry and study only the world line of $\mathfrak{Q}_1$, which coincides in part with the closed timelike curve. $\mathfrak{Q}_1$ thereby becomes a second qubit $\mathfrak{Q}_2$ at an earlier time and in a separate location, represented by a triple $\hat{\boldsymbol{q}}_2(t)$ that conform to the Pauli algebra and are continuous with the $\hat{\boldsymbol{q}}_1(t)$.

The interaction between $\mathfrak{Q}_1$ and $\mathfrak{Q}_2$ can be modelled as their being part of a quantum computational network and passing through a general quantum gate **G** for 1 unit of time. $\mathfrak{Q}_1$

then travels back in time on the closed timelike curve, which in the network-description is represented as its experiencing a negative time-delay of $-1$ unit (figure 1.b).

In the Heisenberg picture, kinematic consistency requires the descriptors of the older qubit $\mathfrak{Q}_2$ immediately after gate **G** equal the descriptors of the younger qubit $\mathfrak{Q}_1$ before participating in **G**, *i.e.*

$$\hat{\boldsymbol{q}}_2(0) = \hat{\boldsymbol{q}}_1(1). \tag{1.4}$$

But now the kinematic consistency condition (1.4) is incompatible with the orthodox assumption that the descriptors of $\mathfrak{Q}_1$ commute with those of $\mathfrak{Q}_2$. The simplest case of incompatibility occurs when **G** is a unit wire, so that the descriptors of $\mathfrak{Q}_1$ remain unchanged between $t = 0$ and $t = 1$; that is to say that $\hat{\boldsymbol{q}}_1(1) = \hat{\boldsymbol{q}}_1(0)$. From this and (1.4), one finds that the descriptors of $\mathfrak{Q}_1$ and $\mathfrak{Q}_2$ are identical at time $t = 0$, so the descriptors of those qubits cannot all commute at $t = 0$.

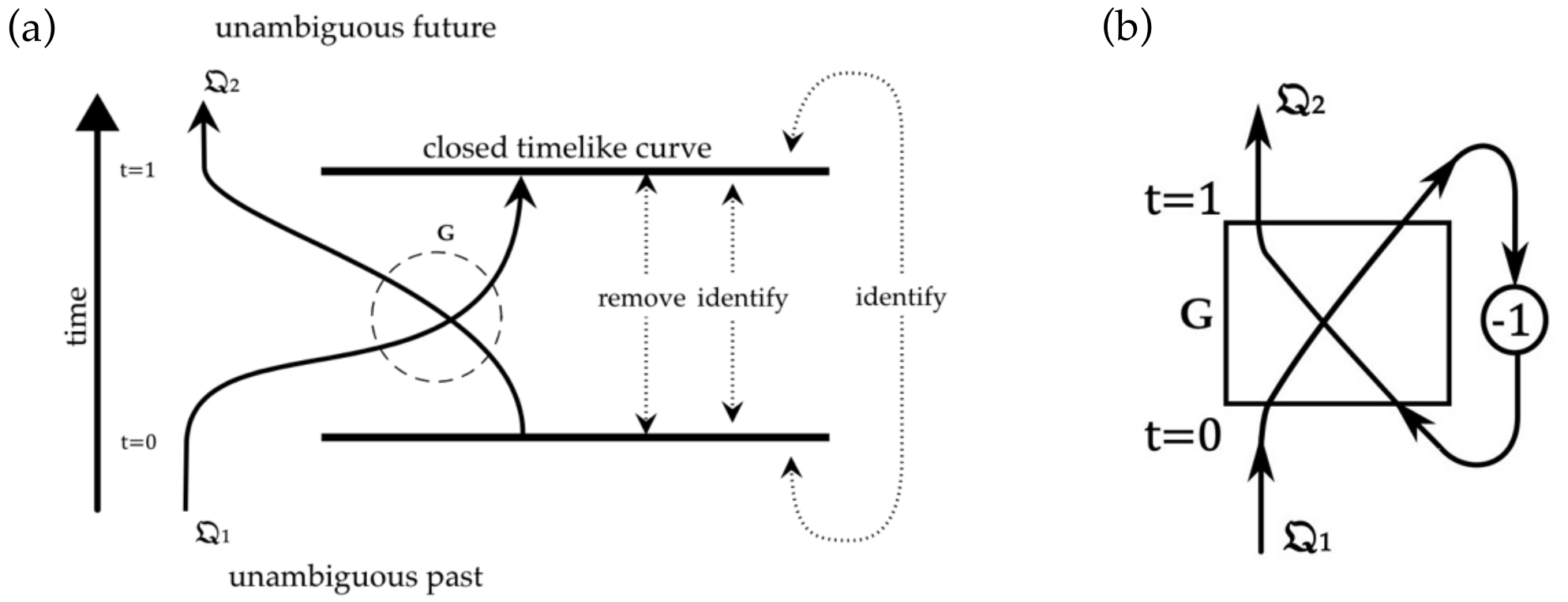

Figure 1. (a) A spacetime diagram depicting the world lines of $\mathfrak{Q}_1$ and $\mathfrak{Q}_2$ and their interacting region G, as well as the unambiguous future and past of the qubit. (b) A network diagram in which $\mathfrak{Q}_1$ enters a two-qubit gate **G**. After leaving the gate, $\mathfrak{Q}_1$ experiences a time-delay of $-1$ before participating in **G** again as the older qubit $\mathfrak{Q}_2$.

There are models of qubits on closed timelike curves that do not encounter this conflict between the kinematic consistency condition (1.4) and the commutation constraint (e.g. Deutsch (1991) and Lloyd *et al.* (2011)). However, those models are formulated in the Schrödinger picture and are therefore not locally realistic (Raymond-Robichaud (2021)). To preserve local realism, qubits on closed timelike curves must treated in the Heisenberg picture, as I have done here; in which case they necessarily violate the commutation constraint.

## 1.2 Quantum field theory

The commutation constraint is also problematic in orthodox quantum field theory. Consider, a scalar quantum field $\hat{\phi}(x)$, where $x$ represents spacetime position. The commutation constraint is

$$[\hat{\phi}(x), \hat{\phi}(y)] = 0$$

when $x$ and $y$ are events on any spacelike hypersurface. Hence, in any compact region $\mathcal{R}$ of space, there are infinitely many mutually commuting observables. So the Hilbert space associated with the field on $\mathcal{R}$ must be infinite dimensional. Yet, as Deutsch (*op. cit.*) and Bao *et al.* (2017) and have argued, that Hilbert space should be finite-dimensional; otherwise, the von Neumann entropy associated with $\mathcal{R}$ would be unbounded, despite it being widely assumed to be bounded (Bekenstein (1973)).

This conflict between the commutation constraint and Bekenstein's bound was one of Deutsch's motivations (*op. cit.*) for discarding that constraint. In the resulting modified quantum field theory – called a 'qubit field theory' by Deutsch – a field's descriptors at spacelike separations do not necessarily commute. Such fields can have finite-dimensional Hilbert spaces. Deutsch conjectured that they can satisfy Bekenstein's formula.

To investigate this, one needs a theory of information and computation for quantum systems that do not commute at spacelike separations. I shall use the discrete version of qubit field theory, which is precisely the theory of the unorthodox qubits of §1.1. When theories of information and measurement have been developed for unorthodox qubits, one can begin to investigate the entropy of qubit fields and determine the truth or falsity of Deutsch's conjecture.

## 2 Unorthodox qubits

I shall call systems that differ from the orthodox theory solely in the omission of constraint (1.1) *unorthodox quantum systems*. In this section, I study the simplest such system, an array $\mathfrak{A}$ of $n$ interacting unorthodox qubits $\mathfrak{Q}_1, \mathfrak{Q}_2, \dots, \mathfrak{Q}_n$.

An unorthodox qubit $\mathfrak{Q}_a$ at time $t$ is represented, like an orthodox one, by a triple of q-number descriptors $\hat{\boldsymbol{q}}_a(t)$ that conforms to the Pauli algebra (1.3). Besides these algebraic constraints, I shall also assume that the descriptors are analytic functions of time so that they can be differentiated arbitrarily often and are equivalent to their Taylor series expansions.

In orthodox quantum theory, one would impose the commutation constraint $[\hat{q}_{ai}(t), \hat{q}_{bj}(t)] = 0$ for all $i, j \in \{x, y, z\}$ when $a \neq b$. In the unorthodox theory, the qubits can also be in states in which $[\hat{q}_{ai}(t), \hat{q}_{bj}(t)] \neq 0$. It is not possible to obtain such a theory by quantising a classical theory, since without the commutation constraint, quantum theory lacks a canonical procedure for promoting c-numbers to q-numbers. So it should also be expected that unorthodox quantum systems will exhibit additional uniquely quantum properties.

## 2.1 Dynamics

Since the descriptors of a qubit $\mathfrak{Q}_a$ always conform to the Pauli algebra (1.3), their time evolution must be unitary. But while in orthodox quantum theory, a single global unitary governs the evolution of all qubits simultaneously, such a unitary would not preserve local realism in the unorthodox case.[4] To preserve both local realism and the algebra of each qubit's descriptors, one must instead allow potentially different unitaries for each qubit. A general qubit $\mathfrak{Q}_a$ is therefore assigned a unitary $U_a(t)$, where the subscript $a$ denotes that this unitary governs the evolution of that qubit alone:

$$\hat{\boldsymbol{q}}_a(t) \stackrel{\text{def}}{=} U_a^\dagger(t)\hat{\boldsymbol{q}}_a(0)U_a(t) \qquad (1 \leq a \leq n), \tag{2.1}$$

where each $U_a(t)$ satisfies

$$U_a^\dagger(t)U_a(t) = \hat{1} \qquad (1 \leq a \leq n), \tag{2.2}$$

with $\hat{1}$ being the unit observable of system $\mathfrak{A}$. So, the evolution of the array $\mathfrak{A}$ is specified by a set of initial conditions $\{\hat{\boldsymbol{q}}_a(0)\}$ and by the $n$ unitaries $\{U_a(t)\}$.

Local realism is preserved by the assumption that the qubits have potentially different unitaries since it allows operations on an individual qubit, or subsystems of qubits, not to affect the rest of the array. Such interactions can be thought of as being caused by 'gates'. So the unitaries of all qubits being acted on by a particular gate are functions only of the descriptors of those qubits.

For example: let the unitary of the qubit $\mathfrak{Q}_a$ be $U_a(t) = e^{-\frac{1}{2}i\omega t\hat{q}_{ax}(0)}$, which rotates that qubit about its $x$-axis by $\omega t$ radians, and let the remaining qubits in the array have unitaries that equal the identity operator, ensuring that those qubits will be unaffected by the rotation performed on $\mathfrak{Q}_a$. From this set of unitaries, and using (2.1), one finds that the descriptors of $\mathfrak{Q}_a$, expressed as functions of the descriptors at $t = 0$, are as follows:

$$\hat{\boldsymbol{q}}_a(t) = \left(\hat{q}_{ax},\ \cos(\omega t)\hat{q}_{ay} + \sin(\omega t)\hat{q}_{az}, \cos(\omega t)\hat{q}_{az} - \sin(\omega t)\hat{q}_{ay}\right), \tag{2.3}$$

while all other qubits in $\mathfrak{A}$ are left unchanged, as required. Here the $t = 0$ labels have been supressed for the sake of brevity. The fact that such operations can be performed, and that therefore the system can have equations of motion in which the qubit's interact only locally, verifies that Einstein's criterion is satisfied.[5]

---

[4] Chao *et al.* (2017) have created a theory of qubits without the commutation constraint, which they named 'overlapping qubits'. Their construction differs from the one presented here in that overlapping qubits evolve according to a global unitary, so the evolution of those qubits is not local.

[5] Because unorthodox qubits may have differing unitaries, there is no Schrödinger picture of unorthodox quantum theory. Nonetheless it is possible to construct a theory of information and measurement for unorthodox qubits – see §§5-6 of this paper.

Each unitary $U_a(t)$ has a generator

$$\hat{H}_a(t) \stackrel{\text{def}}{=} -i\frac{dU_a^\dagger(t)}{dt}U_a(t). \tag{2.4}$$

In analogy with the orthodox theory, I shall call $\hat{H}_a(t)$ the 'Hamiltonian of $\mathfrak{Q}_a$'. It is automatically Hermitian since, by differentiating $U_a^\dagger(t)U_a(t) = \hat{1}$, one finds that $\frac{dU_a^\dagger(t)}{dt}U_a(t)$ is anti-Hermitian.

Unlike in orthodox quantum theory, in which there is only a single Hamiltonian, the array $\mathfrak{A}$ described here has a total of $n$ Hamiltonians – one for each qubit. So each qubit $\mathfrak{Q}_a$ has an equation of motion similar in form to the Heisenberg equation of motion in orthodox quantum theory. By differentiating both sides of (2.1) with respect to *t* and using the definition (2.4), we obtain $\mathfrak{Q}_a$'s equation of motion:

$$\frac{d\hat{\boldsymbol{q}}_a(t)}{dt} = i\big[\hat{H}_a(t), \hat{\boldsymbol{q}}_a(t)\big]. \tag{2.5}$$

Note, from (2.5), that $\hat{H}_a(t)$ is not necessarily a constant of the motion, as its time-derivatives depend on the expressions for the other Hamiltonians $\{\hat{H}_a(t)\}$ which typically change with time. They are not energies.

A Hamiltonian such as $\hat{H}_a(t)$ does not necessarily commute with itself at different times. So there is a factor-ordering problem that prevents the unitary $U_a(t)$ from being expressed as the exponentiated integral of $-i\hat{H}_a(t)$. To find the general expression of this unitary in terms of its associated Hamiltonian, consider that due to (2.4), $U_a(t)$ and $\hat{H}_a(t)$ necessarily satisfy the differential equation

$$i\frac{dU_a(t)}{dt} = U_a(t)\hat{H}_a(t).$$

The solution to this differential equation, with the initial value $U_a(0) = \hat{1}$, is the following time-ordered product (Bauer *et al.* (2012))

$$U_a(t) = \mathcal{T}\exp\left(-i\int_0^t \hat{H}_a(t')\,dt'\right). \tag{2.6}$$

Manifestly, $\hat{H}_a(t)$ determines the expression for $U_a(t)$, up to an arbitrary phase factor. This makes it possible to specify the dynamics of the qubits either in terms of the unitaries $\{U_a(t)\}$ or in terms of the Hamiltonians $\{\hat{H}_a(t)\}$, as one can switch between these two descriptions by using (2.6) and (2.4).

To ensure that the evolution of an isolated system such as $\mathfrak{A}$ depends exclusively on the system's descriptors, I now require that each Hamiltonian can be expressed algebraically in terms of the descriptors of $\mathfrak{A}$ via addition, multiplication, and scaling (multiplication by real numbers):

$$\hat{H}_a(t) = \hat{H}_a\big(\hat{\boldsymbol{q}}_1(t), \dots, \hat{\boldsymbol{q}}_n(t)\big) \qquad (1 \le a \le n). \qquad (2.7)$$

Hence, Hamiltonians cannot explicitly depend on time, as that would violate the assumption that they are functions of the descriptors only. By specifying a set of Hamiltonians that comply with (2.7), one can deduce what the $n$ unitaries $\{U_a(t)\}$ are via (2.6).

The requirement (2.7) holds for all closed systems. And since any open system can be treated as part of a larger closed system, unorthodox qubits fundamentally have Hamiltonians that are functions of the descriptors only. Therefore, I shall only treat systems of unorthodox qubits that adhere to assumption (2.7) in this section.[6]

## 2.2 Invariants of the array

Since qubits in $\mathfrak{A}$ may have different unitaries, the algebraic relations between qubits will generically vary with time. So, unlike in orthodox quantum theory, not all algebraic relations of the qubits are characteristic of the array. In the orthodox theory, a system's time-invariant algebraic properties explain why a system retains its identity despite changing with time. Since those algebraic relations are no longer constants of the motion, in what sense is $\mathfrak{A}$ the same system over time?[7]

This conceptual difficulty would be resolved if one could identify a constitutive (*i.e.* time-invariant) property of $\mathfrak{A}$ that serves to distinguish it from a different array of unorthodox qubits. To that end, consider the algebraic closer (*i.e.* the set of all things derivable by addition, multiplication, and scaling by a real number) of the $3n$ descriptors in $\{\hat{q}_{aj}(t)\}$.[8] I shall denote the algebraic closer of the descriptors as $\boldsymbol{O}(t)$, and it characterises the array.

Although $\boldsymbol{O}(t)$ appears to be a function of time, it is a constant of the motion – *i.e.*, $\boldsymbol{O}(t) = \boldsymbol{O}(0)$ for all times $t$. To prove this, I shall first show that $\boldsymbol{O}(t) \subseteq \boldsymbol{O}(0)$. By definition, any element in $\boldsymbol{O}(t)$ can be expressed algebraically in terms of the descriptors $\{\hat{q}_{aj}(t)\}$, and these descriptors at time $t$ are automatically elements of $\boldsymbol{O}(0)$. That is so because of the assumption that the descriptors are analytic in $t$, and therefore any $\hat{q}_{aj}(t)$ in $\{\hat{q}_{aj}(t)\}$ can be Taylor-expanded around $t = 0$ as follows:

[6] Open systems are defined as those that violate (2.7).

[7] Karl Popper called this *the problem of change*, which he formulated as: "How can a thing change, without losing its identity?" (Popper (1963), p. 192).

[8] $\{\hat{q}_{aj}(t)\}$ denotes the set of all descriptors for the values of $a \in \{1, \dots, n\}$ and $j \in \{x, y, z\}$.

$$\hat{q}_{aj}(t) = \sum_{k=0}^{\infty} \left(\frac{t^k}{n!}\right) \frac{d^k \hat{q}_{aj}(0)}{dt^k}. \tag{2.8}$$

$\hat{q}_{aj}(t)$ is an element of $\boldsymbol{\mathcal{O}}(0)$ because each term in the Taylor expansion is in $\boldsymbol{\mathcal{O}}(0)$. For instance, the first term is $\hat{q}_{aj}(0)$, which is evidently in $\boldsymbol{\mathcal{O}}(0)$. One can readily verify that the second term in the expansion $\frac{d\hat{q}_{aj}(0)}{dt}$ belongs to $\boldsymbol{\mathcal{O}}(0)$ as well by using (2.7) and (2.5) to obtain

$$\frac{d\hat{q}_{aj}(0)}{dt} = i\left[\hat{H}\left(\hat{\boldsymbol{q}}_1(0), \ldots, \hat{\boldsymbol{q}}_n(0)\right), \hat{q}_{aj}(0)\right], \tag{2.9}$$

where the right side of (2.9) is manifestly in $\boldsymbol{\mathcal{O}}(0)$. From (2.9), one can show that the second-order derivatives of the $\{\hat{q}_{aj}(t)\}$ at time $t = 0$ are in $\boldsymbol{\mathcal{O}}(0)$: one takes the derivative of both sides of (2.5) and evaluates the expression at $t = 0$. Thus one finds that the second-order derivative $\frac{d^2\hat{q}_{aj}(0)}{dt^2}$ is a function of both the $\{\hat{q}_{aj}(0)\}$ and the first-order derivatives $\left\{\frac{d\hat{q}_{aj}(0)}{dt}\right\}$. But these first-order derivatives have already been shown to belong to $\boldsymbol{\mathcal{O}}(0)$ in (2.9), and therefore, the second-order derivatives must be functions solely of the $\{\hat{q}_{aj}(0)\}$ – and so on. Hence each of the descriptors in $\{\hat{q}_{aj}(t)\}$ is an element of $\boldsymbol{\mathcal{O}}(0)$.

That is to say, $\boldsymbol{\mathcal{O}}(t) \subseteq \boldsymbol{\mathcal{O}}(0)$. The reverse is also true: each of the $\{\hat{q}_{aj}(0)\}$ can similarly be expressed through a Taylor expansion as an element of $\boldsymbol{\mathcal{O}}(t)$. Thus, $\boldsymbol{\mathcal{O}}(t) = \boldsymbol{\mathcal{O}}(0)$ for any time $t$, implying that $\boldsymbol{\mathcal{O}}(t)$ is a constant of the motion, as promised. The time label of $\boldsymbol{\mathcal{O}}(t)$ is, therefore, superfluous and I will henceforth denote the vector space simply as $\boldsymbol{\mathcal{O}}$.

The algebraic relations among qubits in the array typically change with time, yet the set $\boldsymbol{\mathcal{O}}$ of q-numbers that can be generated algebraically from those qubits' descriptors is invariant. So, an assignment of members of $\boldsymbol{\mathcal{O}}$ to each qubit (with the right Pauli algebra (1.3)) suffices to characterise the array $\mathfrak{A}$ in all its possible states.

## 2.3 Hilbert space

Because $\boldsymbol{\mathcal{O}}$ is a constant of the motion, it is associated with a constant Hilbert space $\boldsymbol{H}$ of dimension $N$, and the elements of $\boldsymbol{\mathcal{O}}$ have a representation on $\boldsymbol{H}$ in terms of $N$-dimensional Hermitian matrices. Similarly, when an individual qubit $\mathfrak{Q}_a$ is considered in isolation, the descriptors of $\mathfrak{Q}_a$ span a subspace of $\boldsymbol{\mathcal{O}}$, namely

$$\boldsymbol{\mathcal{O}}_a(t) = \mathrm{Span}\{\hat{q}_{ax}(t),\ \hat{q}_{ay}(t),\ \hat{q}_{az}(t)\},$$

and this subspace is associated with a 2-dimensional Hilbert space $\boldsymbol{H}_a$.

The subspace $\boldsymbol{\mathcal{O}}_a(t)$ has a complement denoted $\overline{\boldsymbol{\mathcal{O}}}_a(t)$ (Deutsch (*op. cit.*)), which consists of all elements of $\boldsymbol{\mathcal{O}}$ that commute with all elements in $\boldsymbol{\mathcal{O}}_a(t)$. Consequently when two qubits $\mathfrak{Q}_a$ and $\mathfrak{Q}_b$ have mutually commuting descriptors, the descriptors of $\mathfrak{Q}_b$ must be elements of $\bar{\boldsymbol{\mathcal{O}}}_a(t)$, and thus the intersection of their associated subspaces is the empty set:

$$\boldsymbol{\mathcal{O}}_a(t) \cap \boldsymbol{\mathcal{O}}_b(t) = \emptyset.$$

Since there is no overlap between these $\boldsymbol{\mathcal{O}}_a(t)$ and $\boldsymbol{\mathcal{O}}_b(t)$, the descriptors of $\mathfrak{Q}_a$ and $\mathfrak{Q}_b$ are q-numbers on distinct the Hilbert spaces $\boldsymbol{H}_a$ and $\boldsymbol{H}_b$, and so this two-qubit composite system has a Hilbert space that is equal to the tensor product $\boldsymbol{H}_a \otimes \boldsymbol{H}_b$.

But when the descriptors of $\mathfrak{Q}_a$ do *not* commute with those of $\mathfrak{Q}_b$, the intersection of the subspaces $\boldsymbol{\mathcal{O}}_a(t)$ and $\boldsymbol{\mathcal{O}}_b(t)$ is not necessarily empty. Thus, a composite system of unorthodox qubits has a Hilbert space $\boldsymbol{H}$ that is not necessarily a tensor product of the Hilbert spaces of its subsystems. An example of this is a system consisting of two qubits $\mathfrak{Q}_a$ and $\mathfrak{Q}_b$ with the property that

$$\boldsymbol{\mathcal{O}}_a(t) = \boldsymbol{\mathcal{O}}_b(t). \qquad (2.10)$$

(2.10) implies that the descriptors of $\mathfrak{Q}_a$ can be expressed as linear combinations of the descriptors of $\mathfrak{Q}_b$ and vice versa. Therefore, the descriptors of both $\mathfrak{Q}_a$ and $\mathfrak{Q}_b$ are q-numbers on the same 2-dimensional Hilbert space so that the Hilbert space of the composite system is also 2-dimensional, in contrast to the 4-dimensional Hilbert space of two orthodox qubits. When a pair of qubits has property (2.10), they are *maximally non-commuting* ('maximal' because with these commutation relations of the qubits, the dimension of their Hilbert space cannot have a smaller dimension).[9]

Consider an array of $n$ unorthodox qubits with an $N$-dimensional Hilbert space. What could $N$ be? Firstly, the $N$ is greatest when separate qubits in the array have mutually commuting descriptors; in which case $N = 2^n$. But, because the descriptors of separate qubits do not necessarily commute, $N$ is in general much smaller than that – *e.g.* if all $n$ unorthodox qubits maximally non-commute, then $N = 2$.

When $N$ is comparatively small, fewer states are available to the unorthodox qubits than in the orthodox case. Consider, for example, the case of $n$ maximally non-commuting qubits so that $N = 2$. All the array's descriptors can then be represented as 2-dimensional matrices so that each descriptor is a linear combination of, for instance, the Pauli matrices. The most general linear combination of the Pauli matrices that preserves the Pauli algebra is a rigid rotation of those matrices. So, at some general time $t$, the descriptors of an arbitrary qubit in this particular array can be expressed as

$$\hat{\boldsymbol{q}}_a(t) = \hat{R}\big(\theta_a(t), \phi_a(t), \psi_a(t)\big)\hat{\boldsymbol{\sigma}} \qquad (1 \le a \le n). \tag{2.11}$$

---

[9] Notably, this relation is transitive: if $\mathfrak{Q}_a$ maximally non-commutes with $\mathfrak{Q}_b$, and $\mathfrak{Q}_b$ maximally non-commutes with $\mathfrak{Q}_c$, this it implies that $\boldsymbol{\mathcal{O}}_a(t) = \boldsymbol{\mathcal{O}}_b(t) = \boldsymbol{\mathcal{O}}_c(t)$, and hence the qubits $\mathfrak{Q}_a$ and $\mathfrak{Q}_c$ maximally non-commute with one another, too. As a result, a collection of non-commuting qubits of any size must have descriptors that can be expressed in terms of the descriptors of the other qubits, and so such a collection of maximally non-commuting has a 2-dimensional Hilbert space.

Here $\hat{R}$ is a $3 \times 3$ rotation matrix that rotates the triple $\hat{\boldsymbol{\sigma}} \stackrel{\text{def}}{=} \big(\hat{\sigma}_x, \hat{\sigma}_y, \hat{\sigma}_z\big)$ about its $x$-, $y$-, and $z$-axis by the respective angles $\theta_a(t)$, $\phi_a(t)$, and $\psi_a(t)$, which are real numbers in the interval $[0,2\pi)$.

Because of (2.11), the descriptors of each qubit contain at most three c-number parameters, and thus one needs no more than $3n$ parameters to fully specify the $n$ qubit array. To similarly specify the descriptors of an array of $n$ orthodox qubits requires exactly $2^{2n} - 1$ real-valued parameters (Bédard (*op. cit.*)). Moreover, since the descriptors represent the state of the array, exponentially fewer states are available to maximally non-commuting qubits than to the same number of orthodox qubits. This suggests that less information can be stored in qubits that violate the commutation constraint, which was one of the motivations for dropping that constraint, as discussed in §1.2.

## 2.4 Heisenberg state

$\mathfrak{A}$ is fully specified physically by its time-dependent descriptors (2.1) and constant Heisenberg state $|\Psi\rangle$, where the latter is a ray in $\mathfrak{A}$'s Hilbert space $\boldsymbol{H}$. The Heisenberg state is necessary for making predictions, since the expectation values of the descriptors depend on it. As a shorthand, I denote the expectation value of an arbitrary time-dependent descriptor $\hat{A}(t)$ of $\mathfrak{N}$ as

$$\langle \hat{A}(t) \rangle \stackrel{\text{def}}{=} \langle \Psi | \hat{A}(t) | \Psi \rangle.$$

Although $\langle \hat{A}(t) \rangle$ is suggestively named the 'expectation value of $\hat{A}(t)$', it is not evident from the analysis thus far that $\langle \hat{A}(t) \rangle$ must correspond to a measurable quantity. That claim requires a theory of measurement and computation for unorthodox qubits, an outline of which is provided in §6 below.

A descriptor $\hat{A}(t)$ is said to be *sharp* if $\langle \hat{A}(t) \rangle^2 = \langle \hat{A}(t)^2 \rangle$; in which case the Heisenberg state is an eigenstate of the descriptor, and $\langle \hat{A}(t) \rangle$ is one of the eigenvalues of $\hat{A}(t)$. To fix the Heisenberg state of an arbitrary array of unorthodox qubits, I assume, without loss of any relevant generality, that the initial ($t = 0$) $z$-observables of all qubits are sharp with value 1:

$$\langle \hat{q}_{az}(0) \rangle = 1 \qquad (1 \le a \le n). \tag{2.12}$$

This implies that, up to an irrelevant phase factor, the Heisenberg state $|\Psi\rangle$ is equivalent to the eigenstate of the descriptors $\{\hat{q}_{az}(0)\}$ with eigenvalue 1. And since the Heisenberg 'state' is a fixed constant, no information about the state of $\mathfrak{A}$ can be contained by $|\Psi\rangle$; all such information about the state of the array must instead reside in the time-dependent descriptors $\{\hat{\boldsymbol{q}}_a(t)\}$. Incidentally, (2.12) also entails the $\{\hat{q}_{az}(0)\}$ must mutually commute with each other. This loses no generality since the constraint can be satisfied even when all the qubits in the array maximally non-commute (see for example §3).

## 3 A model theory

Consider two interacting unorthodox qubits $\mathfrak{Q}_1$ and $\mathfrak{Q}_2$; at a general time $t$. They are represented by the descriptors

$$\left.\begin{aligned}\hat{\boldsymbol{q}}_1(t) &= \big(\hat{q}_{1x}(t),\ \hat{q}_{1y}(t),\ \hat{q}_{1z}(t)\big)\\ \hat{\boldsymbol{q}}_2(t) &= \big(\hat{q}_{2x}(t),\ \hat{q}_{2y}(t),\ \hat{q}_{2z}(t)\big)\end{aligned}\right\}.$$

Each triple of descriptors conforms to the Pauli algebra, but to fully specify the algebraic relations of the descriptors, I must also define the commutation relations between the two qubits at, for instance, the initial time $t = 0$. I will consider the case where initially the qubits have identical descriptors, *i.e.* $\hat{\boldsymbol{q}}_a(0) = \hat{\boldsymbol{q}}_b(0)$. This implies that the qubits have the following initial commutation-relation $\hat{q}_{1i}(0)\hat{q}_{2j}(0) = \hat{1}\delta_{ij} + i\epsilon_{ij}{}^{k}\hat{q}_{1k}(0)$ for $i, j, k \in \{x, y, z\}$. Therefore, the initial descriptors of $\mathfrak{Q}_a$ and $\mathfrak{Q}_b$ are represented by the Pauli matrices[10]

$$\left.\begin{aligned}\hat{\boldsymbol{q}}_1(0) &= \left(\hat{\sigma}_x,\ \hat{\sigma}_y,\ \hat{\sigma}_z\right)\\ \hat{\boldsymbol{q}}_2(0) &= \left(\hat{\sigma}_x,\ \hat{\sigma}_y,\ \hat{\sigma}_z\right)\end{aligned}\right\}. \tag{3.1}$$

Due to these algebraic relations of the qubits, the model's vector space of Hermitian q-numbers, denoted $\boldsymbol{O}_M$, is spanned by the basis elements $\hat{\sigma}_x, \hat{\sigma}_y, \hat{\sigma}_z$, and $\hat{1}_2$, where $\hat{1}_2$ is the $2 \times 2$ identity matrix. The matrices in $\boldsymbol{O}_M$ live on a 2-dimensional Hilbert space $\boldsymbol{H}_M$. And since the Hilbert space $\boldsymbol{H}_M$ is 2-dimensional, $\mathfrak{Q}_a$ and $\mathfrak{Q}_b$ maximally non-commute.

Maximally non-commuting qubits such as $\mathfrak{Q}_a$ and $\mathfrak{Q}_b$ have special properties. For instance, at time $t = 0$, the qubits $\mathfrak{Q}_a$ and $\mathfrak{Q}_b$ are symmetric under an exchange of their labels because the initial descriptors $\hat{\boldsymbol{q}}_a(0)$ and $\hat{\boldsymbol{q}}_b(0)$ are identical, as shown in (3.1). Another unique feature of maximally non-commuting qubits is that the anticommutator of their descriptors must be a multiple of the unit observable – *i.e.* at an arbitrary time $t$, the descriptors of $\mathfrak{Q}_a$ and $\mathfrak{Q}_b$ have the property that

$$\left\{\hat{q}_{1i}(t), \hat{q}_{2j}(t)\right\} = \mathrm{Tr}\left(\hat{q}_{1i}(t)\hat{q}_{2j}(t)\right)\hat{1} \qquad \text{(for all } i, j \in \{x, y, z\}\text{)}. \tag{3.2}$$

[10] Despite the utility of the matrix representation (3.1), it is merely a way of expressing the descriptors' algebra. One could adopt a different matrix representation or even to do without such a representation altogether, albeit at the cost of making calculations less transparent and more difficult to perform.

One can prove identity (3.2) by noting that the descriptors $\hat{q}_{1i}(t)$ and $\hat{q}_{2j}(t)$ are linear combinations of the anticommuting Pauli matrices and that the Pauli matrices satisfy the identity $\mathrm{Tr}(\hat{\sigma}_i\hat{\sigma}_j) = 2\delta_{ij}$ for $i,j \in \{x,y,z\}$.

I now turn to the model's dynamics, for which the qubits' Hamiltonians must be specified. In this model theory, the qubits' Hamiltonians are defined to be

$$\left.\begin{aligned} \hat{H}_1(\hat{\boldsymbol{q}}_1(t), \hat{\boldsymbol{q}}_2(t)) &= \frac{1}{4}\{\{\hat{q}_{1z}(t), \hat{q}_{2z}(t)\}, \hat{q}_{1x}(t)\} \\ \hat{H}_2(\hat{\boldsymbol{q}}_1(t), \hat{\boldsymbol{q}}_2(t)) &= 0 \end{aligned}\right\}. \tag{3.3}$$

Because the Hamiltonian of $\mathfrak{Q}_b$ is equal to zero, this qubit is stationary. Determining the evolution of $\mathfrak{Q}_a$ is more complicated, but notably its $x$-observable $\hat{q}_{1x}(t)$ is stationary since it commutes with $\hat{H}_1(\hat{\boldsymbol{q}}_1(t), \hat{\boldsymbol{q}}_2(t))$ at all times $t$. Thus, by using (2.6), (3.2) and (3.3), I obtain that

$$\left.\begin{aligned} U_1(t) &= e^{-\frac{i}{2}\alpha(t)\hat{q}_{1x}(0)} \\ U_2(t) &= \hat{1} \end{aligned}\right\},$$

where, to simplify the expression, I introduced the variable

$$\alpha(t) \stackrel{\text{def}}{=} \int_0^t \mathrm{Tr}(\hat{q}_{2z}(0)\hat{q}_{1z}(t'))dt'. \tag{3.4}$$

The unitary $U_1(t)$ represents a rotation about $\mathfrak{Q}_1$'s $x$-axis by $\alpha(t)$ radians, while $\mathfrak{Q}_2$ remains stationary:

$$\left.\begin{aligned} \hat{\boldsymbol{q}}_1(t) &= (\hat{\sigma}_x,\ \cos(\alpha(t))\hat{\sigma}_y + \sin(\alpha(t))\hat{\sigma}_z,\ \cos(\alpha(t))\hat{\sigma}_z - \sin(\alpha(t))\hat{\sigma}_y) \\ \hat{\boldsymbol{q}}_2(t) &= (\hat{\sigma}_x,\ \hat{\sigma}_y,\ \hat{\sigma}_z) \end{aligned}\right\}. \tag{3.5}$$

One can find an exact expression for $\alpha(t)$ by taking the derivative of both sides of definition (3.4) to deduce that $\mathrm{Tr}(\hat{q}_{2z}(0)\hat{q}_{1z}(t)) = \frac{d\alpha(t)}{dt}$. By using that fact and (3.5), we have the self-consistency condition

$$\frac{d\alpha(t)}{dt} = 2\cos(\alpha(t)).$$

Its solution is $\alpha(t) = 2\arctan(\tanh(c+t))$, where $c$ is a real-valued constant that must vanish identically due to the boundary condition $\alpha(0) = 0$ (which follows from the definition (3.4)).

The Heisenberg state is fixed so that the initial $z$-observables of both qubits are sharp with value 1, implying that the Heisenberg state of this model theory, denoted $|\Psi\rangle$, is the eigenstate of $\hat{\sigma}_z$ with eigenvalue 1. From this and (3.5), one can compute the expectation values of the descriptors

$$\left.\begin{aligned}\langle\hat{\boldsymbol{q}}_1(t)\rangle &= \big(0,\ \ \sin\big(\alpha(t)\big),\ \ \cos\big(\alpha(t)\big)\big)\\ \langle\hat{\boldsymbol{q}}_2(t)\rangle &= (0,\ \ 0,\ \ 1)\end{aligned}\right\}.$$

Although the observables of individual qubits are analogous to their orthodox counterparts, the observables of the composite system are not straightforwardly defined because of a factor-ordering problem. Namely, the products of descriptors of separate qubits are not necessarily Hermitian, so they cannot directly represent observables. To obtain the full set of observables of the composite system, one requires a theory of measurement for unorthodox qubits, which does not yet exist in full.[11] However, the foundations for a theory measurement for unorthodox qubits are described in §6.

# 4 Qubit field theory, Bekenstein's bound, and unorthodox qubits

A qubit field is a hypothetical quantum field with descriptors that do not necessarily commute at spacelike separations (Deutsch (*op. cit.*)). At an individual spacetime event, the field's descriptors conform to the Pauli algebra and so are locally identical to an orthodox qubit. Yet, a qubit field cannot be treated as the continuum limit of an array of orthodox qubits since the descriptors of such an array would, by definition, adhere to the commutation constraint. A qubit field should instead be treated as the continuum limit of an array of unorthodox qubits, and consequently studying unorthodox qubits will also inform us about the possible interactions and dynamics of qubit fields.

Because a qubit field does not conform to the commutation constraint, its Hilbert space can be finite-dimensional. And if Deutsch's conjecture is true – *i.e.,* if unorthodox systems with a finite-dimensional Hilbert space have bounded entropy – then qubit fields satisfy Bekenstein's bound, which would resolve an important discrepancy between Bekenstein's bound and quantum field theory. What is currently known about unorthodox qubits seems to suggest that Deutsch's conjecture is true. For example, as discussed in §2.3, the descriptors of an array of maximally non-commuting qubits contain exponentially fewer parameters than the descriptors of an array of orthodox qubits. So perhaps under suitable circumstances (*i.e.* black holes) it would not be possible to store the same amount of information in unorthodox qubits as one can in the orthodox qubits, resulting in a smaller upper-bound on the entropy of an array of unorthodox qubits.

[11] A general notion of measurement and observables – *i.e.* one that is detached from the details of orthodox quantum theory – has been developed by Deutsch & Marletto (2015) through constructor theory. I suggest that their constructor theoretic notion of an observable should be used to define the complete set of observables for unorthodox qubits, which is a problem I shall leave open for future research.

However, since the entropy bound is a measure of the information contained in a system, there must first be a theory of information and measurement for unorthodox qubits before the entropy of such qubits (and qubit fields) can be determined. In the following sections I will provide the foundations of such a theory and demonstrate that unorthodox qubits can instantiate classical information. Because of the above-described connection between unorthodox qubits and qubit fields, the theory of measurement and computation presented in §§5-6 also extends to qubit fields.

# 5 Computational networks

Having treated arrays of unorthodox qubits, in which qubits interact continuously according to some fixed set of Hamiltonians, I shall now consider a *network* $\mathfrak{N}$ consisting of $n$ interacting unorthodox qubits. In a network, qubits interact during periods of fixed duration, which is conventionally assumed to be 1 unit of time, so that the state of the network need only be evaluated at integer times. Between two integer times $t$ and $t+1$, the network's descriptors are smooth functions and perform some general gate **G**, which is implemented by a characteristic set of Hamiltonians $\{\hat{H}_{\mathrm{G},a}(\hat{\boldsymbol{q}}_1(t),\dots,\hat{\boldsymbol{q}}_n(t))\}$ that define the effect of **G** on $\mathfrak{N}$. The set of Hamiltonians are smoothly switched off at the final time $t+1$ and $\mathfrak{N}$ may perform a different gate during the subsequent time step.

That the network's Hamiltonians change at integer times hints at an outside influence on the network's behaviour. That can happen because a network is not an isolated system, as qubits are made to perform gates by an environment (such as an experimenter) whose details are not specified by the network model. However, the environment's only effect on the network is that it changes the qubits' Hamiltonians at integer times. And the way in which this change is brought about by the environment is irrelevant to the dynamics of the network, so the details of the environment can be abstracted away in the network model.

I shall assume that during the period between two integer times $t$ and $t+1$, each descriptor of the network is equal to its Taylor expansion on that interval $[t,t+1)$, so the network is equivalent to the continuously evolving arrays described in earlier sections. This makes the non-analyticity of the descriptors at integer times unimportant, guaranteeing that, like an array of unorthodox qubits, the network's set of q-numbers $\boldsymbol{O}$ is a constant of the motion and that the network has a Hilbert space $\boldsymbol{H}$ of constant dimension $N$.

Since a gate **G** is characterised by a set of Hamiltonians $\{\hat{H}_{\mathrm{G},a}(\hat{\boldsymbol{q}}_1(t),\dots,\hat{\boldsymbol{q}}_n(t))\}$, one can use (2.6) and integrate from $t$ to $t+1$ to produce a set of characteristic unitaries, denoted as $\{U_{\mathrm{G},a}(\hat{\boldsymbol{q}}_1(t),\dots,\hat{\boldsymbol{q}}_n(t))\}$, that similarly define the effect of gate **G**. Each of these unitaries is represented as a function of the $\{\hat{\boldsymbol{q}}_a(t)\}$ which is possible since, using the results of §2.2, one can express any element in $\mathcal{O}$ as function of those descriptors. In terms of the unitaries, the effect of gate **G** on a qubit $\mathfrak{Q}_a$ is

$$\hat{\boldsymbol{q}}_a(t+1) = U^\dagger_{\mathrm{G},a}\big(\hat{\boldsymbol{q}}_1(t),\dots,\hat{\boldsymbol{q}}_n(t)\big)\hat{\boldsymbol{q}}_a(t)U_{\mathrm{G},a}\big(\hat{\boldsymbol{q}}_1(t),\dots,\hat{\boldsymbol{q}}_n(t)\big). \tag{5.1}$$

Unlike in the orthodox case, not every set of unitaries represents a possible gate. For instance, if the descriptors $\hat{\boldsymbol{q}}_a(t)$ and $\hat{\boldsymbol{q}}_b(t)$ of $\mathfrak{Q}_a$ and $\mathfrak{Q}_b$ commute at time $t$, then without further constraints on their unitaries, one could enact the **swap** gate on $\mathfrak{Q}_a$ and simultaneously subject $\mathfrak{Q}_b$ to a unit wire.[12] The effect of this would be that $\hat{\boldsymbol{q}}_a(t+1) = \hat{\boldsymbol{q}}_b(t)$ and $\hat{\boldsymbol{q}}_b(t+1) = \hat{\boldsymbol{q}}_b(t)$, which reduces the dimension of $\mathcal{O}$, contrary to the results of §2.2.

This asymmetry between unitaries and Hamiltonians happens because the unitaries do not specify the intermediate states of the qubits. The Hamiltonians $\{\hat{H}_{\mathrm{G},a}\big(\hat{\boldsymbol{q}}_1(t),\dots,\hat{\boldsymbol{q}}_n(t)\big)\}$ do, and not all unitaries are derivable from Hamiltonians. Since the Hamiltonians give the fundamental definition of a gate, I shall characterise gates in terms of their Hamiltonians.

# 6 Classical computation

I now turn to the set of gates that non-orthodox networks can perform. One salient question is, can they implement classical gates, and if so, which ones? As it turns out, $\mathfrak{N}$ can enact all classical gates. That is so because, as I shall demonstrate, a network of even maximally non-commuting unorthodox qubits can perform a **ccnot** gate – a gate that is universal for classical reversible computation. But before I can define this gate, I have to clarify certain technical issues, such as what a 'classical' gate means.

## 6.1 The unorthodox not gate

An example of a classical gate is the single-qubit **not** operation performed on qubit $\mathfrak{Q}_a$, which toggles the value of that qubit's $z$-observable while it leaves all other qubits in $\mathfrak{N}$ unchanged. This **not** gate is enacted during the period starting at $t$ and ending at $t+1$ if, during that period, the Hamiltonian of $\mathfrak{Q}_a$ is

$$\hat{H}_{\mathbf{not},a}\big(\hat{\boldsymbol{q}}_a(t)\big) = \frac{\pi}{2}\hat{q}_{ax}(t), \tag{6.1}$$

while all other Hamiltonians of $\mathfrak{N}$ are equal to zero, such that the other qubits are not affected by this gate. When the Hamiltonian (6.1) is enacted for 1 unit of time, it represents a rotation of $\mathfrak{Q}_a$ by $\pi$ radians about this qubit's $x$-axis (see for example (2.3)). Thus, by using (2.1), (2.6) and (6.1), one finds that the descriptors of $\mathfrak{Q}_a$ at time $t+1$, expressed as functions of the descriptors at $t$, are

$$\hat{\boldsymbol{q}}_a(t+1) = \left(\hat{q}_{ax}(t),\ \ -\hat{q}_{ay}(t),\ \ -\hat{q}_{az}(t)\right),$$

while all other qubits of $\mathfrak{N}$ remain unchanged. That the **not** gate toggles the $z$-observable of $\mathfrak{Q}_a$ is made explicit by the algebraic relation $\hat{q}_{az}(t+1) = -\hat{q}_{az}(t)$.

[12] The corresponding unitaries for this operation would be $U_a\big(\hat{\boldsymbol{q}}_a(t),\hat{\boldsymbol{q}}_b(t)\big) = \frac{1}{4}\Big(\hat{1} + \delta^{ij}\hat{q}_{ai}(t)\hat{q}_{bj}(t)\Big)$, which represents a **swap** gate, and $U_b\big(\hat{\boldsymbol{q}}_a(t),\hat{\boldsymbol{q}}_b(t)\big) = \hat{1}$, a unit wire. The product in $U_a\big(\hat{\boldsymbol{q}}_a(t),\hat{\boldsymbol{q}}_b(t)\big)$ is well-defined provided that $\hat{\boldsymbol{q}}_a(t)$ and $\hat{\boldsymbol{q}}_b(t)$ commute at the initial time $t$.

The **not** gate is a 'classical gate' because if the input $\hat{q}_{az}(t)$ is sharp, the **not** gate invariably produces a sharp output $\hat{q}_{az}(t+1)$. Hence this $z$-observable behaves identically to how a classical bit would if it were subjected to a **not** operation. More generally, consider the sets of states of qubit $\mathfrak{Q}_a$ for which its $z$-observables are sharp with value 1 and $-1$,

$$\left.\begin{aligned} \check{1}_a &\stackrel{\text{def}}{=} \{\hat{\boldsymbol{q}}_a : \langle\hat{q}_{az}\rangle = 1\} \\ -\check{1}_a &\stackrel{\text{def}}{=} \{\hat{\boldsymbol{q}}_a : \langle\hat{q}_{az}\rangle = -1\} \end{aligned}\right\}. \tag{6.2}$$

The carons in $\check{1}_a$ and $-\check{1}_a$ symbolise that these are sets of states, not numbers or operators, and for the sake of brevity, I have dropped the time labels in (6.2). Following Marletto & Vedral (2020), I shall call sets such as (6.2) *attributes* of a qubit. In terms of attributes, the effect of a **not** gate on $\mathfrak{Q}_a$ can be summarised as $\left\{\check{1}_a \stackrel{\textbf{not}}{\longrightarrow} -\check{1}_a,\ -\check{1}_a \stackrel{\textbf{not}}{\longrightarrow} \check{1}_a\right\}$, where the arrow shows which input attributes are transformed into which output attributes, and where all other qubits are left unchanged.

More generally, a gate **C** will be called 'classical' if it sends input states for which all the qubits' $z$-observables are sharp to output states in which they are all sharp. Thus, under the influence of a classical gate **C**, the sharp $z$-observables of the network will evolve exactly *as if* they were bits in a classical computer. Yet, even in these circumstances, the network as a whole is never classical because it always has some observables that are non-sharp, such as for example the $x$- and $y$-observables of the qubits.

## 6.2 The unorthodox cnot gate

A second important classical gate is the **cnot**, which is a two-qubit gate that acts on a *control qubit* $\mathfrak{Q}_a$ and a *target qubit* $\mathfrak{Q}_b$, with descriptors

$$\left.\begin{aligned} \hat{\boldsymbol{q}}_a(t) &= \left(\hat{q}_{ax}(t),\ \hat{q}_{ay}(t),\ \hat{q}_{az}(t)\right) \\ \hat{\boldsymbol{q}}_b(t) &= \left(\hat{q}_{bx}(t),\ \hat{q}_{by}(t),\ \hat{q}_{bz}(t)\right) \end{aligned}\right\}.$$

The characteristic effect of the **cnot** is that it enacts a **not** on the target $\mathfrak{Q}_b$ if the control has attribute $-\check{1}_a$, and leaves the target attribute unchanged if the control has the attribute $\check{1}_a$. Thus the **cnot** performs

$$\left.\begin{array}{ll}(\check{1}_a,\check{1}_b)\xrightarrow{\textbf{cnot}}(\check{1}_a,\check{1}_b), & (-\check{1}_a,\check{1}_b)\xrightarrow{\textbf{cnot}}(\check{1}_a,-\check{1}_b)\\ (\check{1}_a,-\check{1}_b)\xrightarrow{\textbf{cnot}}(\check{1}_a,-\check{1}_b), & (-\check{1}_a,-\check{1}_b)\xrightarrow{\textbf{cnot}}(-\check{1}_a,\check{1}_b)\end{array}\right\}. \tag{6.3}$$

To have a valid **cnot** gate for unorthodox qubits, there should be a set of Hamiltonians that comply with the principles established in §2 and perform the transformation (6.3) regardless of the commutation relations of the qubits. In orthodox quantum theory, the expressions for the **cnot** gate relies crucially on the fact that the descriptors of separate qubits mutually commute – *e.g.* the expression for the Hamiltonian of the orthodox **cnot** gate contains products of descriptors that would not be Hermitian if those descriptors failed to commute. And most straightforward ways of making such products Hermitian will make the gate lose its functionality. For instance, if all products of descriptors were to be replaced by anticommutators, this would ensure that the expression for the **cnot** gate is Hermitian, but because of (3.2), the anticommutators of descriptors reduce to multiples of the identity if the qubits maximally non-commute so the gate would then have a trivial effect and not be functionally equivalent to a **cnot** gate for unorthodox qubits. To obviate this issue, I shall introduce a *reference qubit* $\mathfrak{Q}_c$, which has descriptors

$$\hat{\boldsymbol{q}}_c(t)=\left(\hat{q}_{cx}(t),\ \ \hat{q}_{cy}(t),\ \ \hat{q}_{cz}(t)\right).$$

With the help of such a qubit one can construct a gate that invariably performs transformation (6.3): I impose that the reference and the control qubits maximally non-commute. This allows me to construct an operator with properties similar to those of a projector, namely

$$\vec{P}\left(\hat{q}_{az}(t),\hat{q}_{cz}(t)\right)\stackrel{\text{def}}{=}\frac{1}{4}\left(2-\{\hat{q}_{az}(t),\hat{q}_{cz}(t)\}\right). \tag{6.4}$$

$\vec{P}\left(\hat{q}_{az}(t),\hat{q}_{cz}(t)\right)$ has the following properties:

(i) It commutes with $\hat{q}_{bx}(t)$, regardless of the commutation relations between the target $\mathfrak{Q}_b$ and the control $\mathfrak{Q}_a$.

(ii) It vanishes when the control has attribute $\check{1}_a$, and equals unity when the control has attribute $-\check{1}_a$.

The arrow in $\vec{P}\left(\hat{q}_{az}(t),\hat{q}_{cz}(t)\right)$ denotes that it is a superoperator, meaning that it maps q-numbers to q-numbers. $\vec{P}\left(\hat{q}_{az}(t),\hat{q}_{cz}(t)\right)$ has property (i) because $\mathfrak{Q}_a$ and $\mathfrak{Q}_c$ are assumed to maximally non-commute, so $\vec{P}\left(\hat{q}_{az}(t),\hat{q}_{cz}(t)\right)$ is a multiple of the identity operator, as can be verified using (6.4) and (3.2). Consequently, $\vec{P}\left(\hat{q}_{az}(t),\hat{q}_{cz}(t)\right)$ commutes with any q-number. The superoperator also has property (ii): if the $z$-observables of $\mathfrak{Q}_a$ and $\mathfrak{Q}_c$ are aligned (*i.e.*

$\hat{q}_{az}(t) = \hat{q}_{cz}(t)$), then $\vec{P}(\hat{q}_{az}(t), \hat{q}_{cz}(t))$ is identically zero; whereas if those same descriptors are anti-aligned (*i.e.* $\hat{q}_{az}(t) = -\hat{q}_{cz}(t)$), then $\vec{P}(\hat{q}_{az}(t), \hat{q}_{cz}(t))$ is equal to the unit observable $\hat{1}$.

Notably, because $\mathfrak{Q}_a$ maximally non-commutes with $\mathfrak{Q}_c$, the descriptors $\hat{q}_{az}(t)$ and $\hat{q}_{cz}(t)$ can only be sharp at the same time if those descriptors are identical up to a minus sign. Because of this, $\mathfrak{Q}_c$ can function as *reference* that indicates whether $\mathfrak{Q}_a$ has attribute $\check{1}_a$ or $-\check{1}_a$: without loss of relevant generality, let the reference qubit have the attribute $\check{1}_c$ at the initial time $t$. By preparing the reference qubit in this way, the superoperator $\vec{P}(\hat{q}_{az}(t), \hat{q}_{cz}(t))$ is equal to zero if the control has the attribute $\check{1}_a$, as then the $z$-observables of the control and reference are aligned; and if the control has attribute $-\check{1}_a$, the $z$-observables of the control and the reference are anti-aligned and $\vec{P}(\hat{q}_{az}(t), \hat{q}_{cz}(t)) = \hat{1}$, as required.

Using the superoperator (6.4), one can define an unorthodox **cnot** gate as follows: between time $t$ and $t+1$, let the Hamiltonian of the target qubit $\mathfrak{Q}_b$ is

$$\widehat{H}_{\mathbf{cnot},b}(\hat{\boldsymbol{q}}_a(t), \hat{\boldsymbol{q}}_b(t), \hat{\boldsymbol{q}}_c(t)) \stackrel{\text{def}}{=} \frac{\pi}{2}\hat{q}_{bx}(t)\vec{P}(\hat{q}_{az}(t), \hat{q}_{cz}(t)), \tag{6.5}$$

while the Hamiltonians of the control and reference qubits are equal to zero. (As in previous examples, the latter condition ensures that $\mathfrak{Q}_a$ and $\mathfrak{Q}_c$ remain stationary between $t$ and $t+1$.) Because of property (i), the product of $\vec{P}(\hat{q}_{az}(t), \hat{q}_{cz}(t))$ and $\frac{\pi}{2}\hat{q}_{bx}(t)$ in (6.5) is well-defined irrespective of the commutation relations of $\mathfrak{Q}_a$ and $\mathfrak{Q}_c$. And due to property (ii), the gate performs a **not** on the target if the control has attribute $-\check{1}_a$ and enacts a unit wire if the control has the attribute $\check{1}_a$. The $z$-observable of the control remains unchanged in either case because that qubit is stationary, implying that this gate performs the transformation (6.3) regardless of the commutation relations of those qubits. So, the gate described in (6.5) is precisely a **cnot** gate for the unorthodox qubits $\mathfrak{Q}_a$ and $\mathfrak{Q}_b$. (It is not the only one.)

The reference qubit is necessary for enacting the **cnot** because, when $\mathfrak{Q}_a$ and $\mathfrak{Q}_b$ maximally non-commute, it appears that only the value of $\hat{q}_{az}(t)$ relative to the value of $\hat{q}_{cz}(t)$ can be measured by $\mathfrak{Q}_b$.[13] Consequently, it appears that the **cnot** gate defined in (6.5) is a three-qubit gate rather than a two-qubit gate because it acts on $\mathfrak{Q}_a$, $\mathfrak{Q}_b$ and $\mathfrak{Q}_c$. However, the reference qubit is left unaffected by the gate. I shall also assume that $\mathfrak{Q}_c$ does not interact with any other qubit in the network before or after it participates in the **cnot**, so that its only role in the network is enabling $\mathfrak{Q}_a$ and $\mathfrak{Q}_b$ to perform the **cnot**. In this sense $\mathfrak{Q}_c$ is similar to an ancilla qubit.

[13] Conceivably, there are other sets of Hamiltonians that also enact the transformation (6.3) and do not require the need for a reference qubit. But I conjecture that there aren't.

Reference qubits should be an available resource within unorthodox quantum theory. Such qubits can, for instance, be produced by qubit fields: a qubit field varies smoothly so that the field's descriptors at two sufficiently nearby locations are arbitrarily close to being identical, implying that the field's descriptors at those locations are equivalent to two maximally non-commuting qubits. One of these qubits can therefore be used as a reference for the other. Similarly, closed timelike curves can produce a pair of qubits that have identical descriptors, as explained in §1.1, and are therefore another source of maximally non-commuting qubits.

The **cnot** defined in (6.5) provides the foundations for a theory of measurement. For example, if the initial state of the target qubit is such that $\langle \hat{q}_{bz}(t) \rangle = 1$, then the **cnot** gate imprints the value $\langle \hat{q}_{az}(t) \rangle$ of the control qubit on the $z$-observable of the target since in this situation the effect of the gate is that $\langle \hat{q}_{bz}(t+1) \rangle = \langle \hat{q}_{az}(t) \rangle$. Hence, expectation values such as $\langle \hat{q}_{bz}(t) \rangle$ correspond to measurable quantities.[14] And because of the connection discussed in §4, this theory of measurement and computation provides the basis for an investigation into the entropy of unorthodox systems.

## 6.3 The unorthodox ccnot gate

Having defined a **cnot** gate, one can similarly construct a **ccnot** gate for unorthodox qubits. The **ccnot** is a three-qubit gate that acts on two control qubits, $\mathfrak{Q}_a$ and $\mathfrak{Q}_b$, and a target $\mathfrak{Q}_c$; the gate performs a **not** on the target qubit if and only if the controls have attributes $(-\check{1}_a, -\check{1}_b)$, whereas if the control qubits have attributes $(\check{1}_a, -\check{1}_b)$, $(-\check{1}_a, \check{1}_b)$, or $(\check{1}_a, \check{1}_b)$, the gate does nothing. Just like the **cnot** gate, the control qubits $\mathfrak{Q}_a$ and $\mathfrak{Q}_b$ will be associated with respective reference-qubits $\mathfrak{Q}_d$ and $\mathfrak{Q}_e$, where the control qubits maximally non-commute with their associated reference qubit. Let us also assume, without loss of relevant generality, that at the initial time $t$, the reference qubits have attributes $\check{1}_d$ and $\check{1}_e$. Then, a **ccnot** is implemented between $t$ and $t+1$ if the target qubit's Hamiltonian is

$$\begin{aligned} &\hat{H}_{\mathbf{ccnot},c}\big(\hat{\boldsymbol{q}}_a(t), \hat{\boldsymbol{q}}_b(t), \hat{\boldsymbol{q}}_c(t), \hat{\boldsymbol{q}}_d(t), \hat{\boldsymbol{q}}_e(t)\big) \\ &\qquad \stackrel{\text{def}}{=} \frac{\pi}{2} \hat{q}_{cx}(t) \vec{P}\big(\hat{q}_{az}(t), \hat{q}_{dz}(t)\big) \vec{P}\big(\hat{q}_{bz}(t), \hat{q}_{ez}(t)\big), \end{aligned} \tag{6.6}$$

while the Hamiltonian of the qubits $\mathfrak{Q}_a$, $\mathfrak{Q}_b$, $\mathfrak{Q}_d$, and $\mathfrak{Q}_e$ are each equal to zero. This latter requirement again ensures that only the target qubit changes between $t$ and $t+1$. Due to the term $\vec{P}\big(\hat{q}_{az}(t), \hat{q}_{dz}(t)\big)\vec{P}\big(\hat{q}_{bz}(t), \hat{q}_{ez}(t)\big)$ in (6.6), this Hamiltonian reduces to that of a **not** gate on the target qubit if the control qubits have attributes $(-\check{1}_a, -\check{1}_b)$, while that same term vanishes identically if control qubits have the attributes $(\check{1}_a, -\check{1}_b)$, $(-\check{1}_a, \check{1}_b)$, or $(\check{1}_a, \check{1}_b)$. Hence, unorthodox qubits can perform a **ccnot** gate. Because of the existence of the **ccnot** – a gate that is universal for reversible classical computations – the network $\mathfrak{N}$ can perform any classical reversible computation. This makes it possible for unorthodox qubits to instantiate classical information.

[14] It is not yet clear what the full set of observables consist of. I shall leave this question open for future research.

Classical gates are only a subset of $\mathfrak{N}$'s computational repertoire. For instance, $\mathfrak{N}$ can also enact certain quantum gates, like a $\sqrt{\textbf{not}}$ gate, for which there is no classical counterpart, and which can be performed by enacting the **not** gate for half the standard duration. So, although the full computational repertoire of $\mathfrak{N}$ has yet to be determined, it must have significant overlap with the computational repertoire of orthodox networks. There is also the possibility that $\mathfrak{N}$ can perform entirely novel kinds of computations.

# 7 Closed timelike curves revisited

One of the motivations for dropping the commutation constraint is that this allows one to analyse the behaviour of quantum systems on closed timelike curves in the Heisenberg picture. I shall now investigate a time travel 'paradox' and demonstrate how the paradox is resolved when the commutation constraint is dropped. I shall also show how the Hilbert space of unorthodox qubits can change in dimension due to the presence of closed timelike curves.

## 7.1 The grandfather paradox

Consider a classical bit $x_1$ that can take the values 1 and $-1$. If it encounters a closed timelike curve and travels on it, it becomes an older bit, $x_2 \in \{1, -1\}$ at an earlier time. Before $x_1$ is sent back in time, the bits $x_1$ and $x_2$ participate in the classical network depicted in figure 2. Because of the gates enacted in this classical network, $x_1$ and $x_2$ will have a paradoxical history. The network in figure 2 consist of a **cnot,** of which $x_1$ is the target and $x_2$ the control, followed by a **not** gate performed on $x_1$. Therefore, if the input state of the bits at time $t = 0$ is $(x_1, x_2)$, the network produces the following output state at $t = 2$:

$$(x_1, x_2) \rightarrow (-x_1 x_2, x_2).$$

Kinematic consistency now requires that the older at time $t = 0$ bit is equal to the younger bit at time $t = 2$ so that $x_2 = -x_1 x_2$. Therefore, if $x_1$ initially has value 1, a contradiction ensues: there is no value for $x_2 \in \{1, -1\}$ that can satisfy this consistency condition.

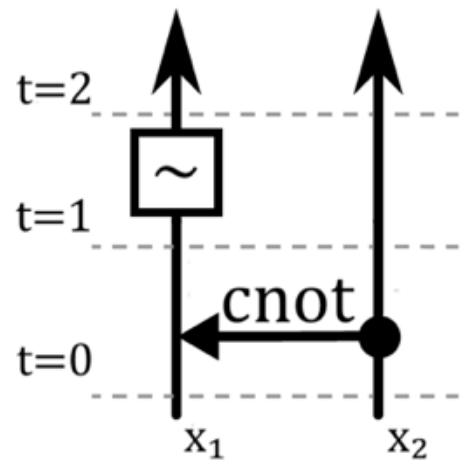

Figure 2. The classical network that the bits $x_1$ and $x_2$ perform.
The network consists of a **cnot** gate, of which $x_1$ is the target and
$x_2$ the control, followed by a **not** on $x_1$.

The situation described above is the essence of the so-called grandfather paradox, in which one observes an older version of oneself who has travelled back in time, but upon seeing this older version, one decides not to travel back in time; thereby preventing the observed history from taking place.

Now let us analyse the grandfather paradox in orthodox quantum theory. Consider a qubit $\mathfrak{Q}_1$, with descriptors $\hat{\boldsymbol{q}}_1(t)$ that conform to the Pauli algebra, that is sent back in time on a closed timelike curve. It becomes a qubit emerging at an earlier time, denoted $\mathfrak{Q}_2$, with descriptors $\hat{\boldsymbol{q}}_2(t)$. Let $\mathfrak{Q}_1$ and $\mathfrak{Q}_2$ interact for 2 units of time, during which they enact the quantum equivalent of the network shown in figure 2, where $x_1$ and $x_2$ respectively correspond to $\mathfrak{Q}_1$ and $\mathfrak{Q}_2$. Furthermore, by analogy with the classical paradox, suppose that $\hat{q}_{1z}(0)$ is sharp with value 1 at the initial time $t = 0$.

In the Heisenberg picture of orthodox quantum theory, the effect of these gates can be summarised as follows:

$$\left.\begin{array}{l}\begin{Bmatrix}\hat{\boldsymbol{q}}_1(0)\\ \hat{\boldsymbol{q}}_2(0)\end{Bmatrix} = \begin{Bmatrix}(\hat{q}_{1x},\ \hat{q}_{1y},\ \hat{q}_{1z})\\ (\hat{q}_{2x},\ \hat{q}_{2y},\ \hat{q}_{2z})\end{Bmatrix}\\ \xrightarrow{\textbf{cnot}}\begin{Bmatrix}\hat{\boldsymbol{q}}_1(1)\\ \hat{\boldsymbol{q}}_2(1)\end{Bmatrix} = \begin{Bmatrix}(\hat{q}_{1x},\ \hat{q}_{1y}\hat{q}_{2x},\ \hat{q}_{1z}\hat{q}_{2x})\\ (\hat{q}_{1x}\hat{q}_{2x},\ \hat{q}_{1x}\hat{q}_{2x},\ \hat{q}_{2x})\end{Bmatrix}\\ \xrightarrow{\textbf{not},1}\begin{Bmatrix}\hat{\boldsymbol{q}}_1(2)\\ \hat{\boldsymbol{q}}_2(2)\end{Bmatrix} = \begin{Bmatrix}(\hat{q}_{1x}, -\hat{q}_{1y}\hat{q}_{2x}, -\hat{q}_{1z}\hat{q}_{2x})\\ (\hat{q}_{1x}\hat{q}_{2x},\ \hat{q}_{1x}\hat{q}_{2x},\ \hat{q}_{2x})\end{Bmatrix}\end{array}\right\}. \tag{7.1}$$

Here the arrows point to the output descriptors produced by the gates, and I have supressed the $t = 0$ labels for the sake of brevity. At the final time $t = 2$, $\mathfrak{Q}_1$ is sent back in time along the closed timelike curve, which is modelled as it experiencing a negative time-delay of $-2$ units of time, after which it is called $\mathfrak{Q}_2$. To ensure kinematic consistency, one must now require that

$$\hat{\boldsymbol{q}}_2(0) = \hat{\boldsymbol{q}}_1(2). \tag{7.2}$$

It follows from (7.2), (7.1) and the Pauli algebra that the descriptors $\hat{\boldsymbol{q}}_2(0)$ cannot all commute with $\hat{\boldsymbol{q}}_1(0)$. For instance, kinematic consistency requires that $\hat{q}_{1x} = \hat{q}_{2x}$, so from the expressions in (7.1), $\hat{q}_{1z}(2) = -\hat{q}_{1z}\hat{q}_{2x} = -i\hat{q}_{1y}$, which clearly violates the requirement that the descriptors be Hermitian. Hence, the Heisenberg picture, in orthodox quantum theory, does not seem to permit a resolution to the grandfather paradox.

However, consistency can be restored if $\mathfrak{Q}_1$ and $\mathfrak{Q}_2$ are unorthodox qubits.[15] Consider, for instance, the case in which $\mathfrak{Q}_1$ and $\mathfrak{Q}_2$ maximally non-commute, so the descriptors of $\mathfrak{Q}_1$ at $t = 0$ could have the matrix representation

$$\hat{\boldsymbol{q}}_1(0) = \left(\hat{\sigma}_x,\ \hat{\sigma}_y,\ \hat{\sigma}_z\right). \tag{7.3}$$

The initial matrix representation of $\mathfrak{Q}_2$ will be a triple of linear combinations of the Pauli matrices, determined by (7.1) and by the gates that act. Assume also that the Heisenberg state is an eigenstate of $\hat{q}_{1z}(0)$ with eigenvalue 1 so that this descriptor is sharp with value 1 at the initial time $t = 0$ like the classical bit $x_1$.

$\mathfrak{Q}_1$ and $\mathfrak{Q}_2$ will undergo the same network as before. However, since $\mathfrak{Q}_1$ and $\mathfrak{Q}_2$ maximally non-commute, an unorthodox **cnot** gate must act on those qubits, and this requires a reference qubit, denoted $\mathfrak{Q}_3$. As has been established in §5, the properties of $\mathfrak{Q}_3$ should be that it maximally non-commute with the control qubit $\mathfrak{Q}_1$; that it remain unaffected by its participation in the network; and that its $z$-observable be sharp value 1. Hence, descriptors for that reference that satisfy these constraints are $\hat{\boldsymbol{q}}_3(0) = \left(\hat{\sigma}_x,\ \hat{\sigma}_y,\ \hat{\sigma}_z\right)$.

An unorthodox **cnot** gate is implemented if the Hamiltonian of $\mathfrak{Q}_1$ is

$$\hat{H}_{\mathbf{cnot},1}\left(\hat{\boldsymbol{q}}_1(t), \hat{\boldsymbol{q}}_2(t), \hat{\boldsymbol{q}}_3(t)\right) = \frac{\pi}{2}\hat{q}_{1x}(t)\vec{P}\left(\hat{q}_{2z}(t), \hat{q}_{3z}(t)\right), \tag{7.4}$$

while the Hamiltonians for $\mathfrak{Q}_2$ and $\mathfrak{Q}_3$ are equal to zero. Furthermore, the descriptors $\hat{q}_{1x}(t)$, $\hat{q}_{2z}(t)$, and $\hat{q}_{3z}(t)$ do not evolve between $t = 0$ and $t = 1$ ($\mathfrak{Q}_2$ and $\mathfrak{Q}_3$ are stationary during this period, and $\hat{q}_{1x}(t)$ commutes with the Hamiltonian of $\mathfrak{Q}_1$). And since the qubits $\mathfrak{Q}_1$, $\mathfrak{Q}_2$ and $\mathfrak{Q}_3$ all maximally non-commute with one another, one can use (3.2) to simplify the expression (7.4) to $\hat{H}_{\mathbf{cnot},1}\left(\hat{\boldsymbol{q}}_1(t), \hat{\boldsymbol{q}}_2(t), \hat{\boldsymbol{q}}_3(t)\right) = \frac{\pi}{8}\hat{q}_{1x}(0)\left(2 - \mathrm{Tr}\left(\hat{q}_{2z}(0)\hat{q}_{3z}(0)\right)\right)$.

[15] In the formalism created by Marletto *et al.* (2019), the state of a qubit on a closed timelike curve is described by a pseudo-density operator Fitzsimons *et al.* (2016), which, unlike the conventional density operator, can describe both spatial and temporal correlations. The approach of Marletto *et al.* and my own presented here should be entirely compatible, and the relationship between these two descriptions might be similar to that between the density-operator formalism and the Heisenberg picture of orthodox quantum theory.

Manifestly, during the period starting at $t = 0$ and ending at $t = 1$, the **cnot** rotates $\mathfrak{Q}_1$ about its $x$-axis by an angle $\phi \stackrel{\text{def}}{=} \frac{\pi}{4}\left(2 - \mathrm{Tr}\left(\hat{q}_{2z}(0)\hat{q}_{3z}(0)\right)\right)$, and so when the qubits stop interacting at time $t = 1$, $\mathfrak{Q}_1$ is in the state

$$\hat{\boldsymbol{q}}_1(1) = \left(\hat{\sigma}_x,\ \hat{\sigma}_y \cos(\phi) + \hat{\sigma}_z \sin(\phi),\ \hat{\sigma}_z \cos(\phi) - \hat{\sigma}_y \sin(\phi)\right), \tag{7.5}$$

while the other descriptors of the network are unchanged. Subsequently, between $t = 1$ and $t = 2$, a **not** gate is performed on $\mathfrak{Q}_1$, which has the following effect on that qubit:

$$\hat{\boldsymbol{q}}_1(2) = \left(\hat{\sigma}_x,\ -\hat{\sigma}_y \cos(\phi) - \hat{\sigma}_z \sin(\phi),\ -\hat{\sigma}_z \cos(\phi) + \hat{\sigma}_y \sin(\phi)\right). \tag{7.6}$$

Because of (7.6) and (7.1), one now has an expression for $\hat{\boldsymbol{q}}_2(0)$. Using this expression and the definition of $\phi$, the self-consistency condition becomes

$$\phi = \frac{\pi}{2}(1 + \cos(\phi)).$$

The unique solution to this self-consistency condition is that $\phi = \frac{\pi}{2}$, so the evolution of $\mathfrak{Q}_1$, $\mathfrak{Q}_2$ and $\mathfrak{Q}_3$ between $t = 0$ and $t = 2$ is:

$$\left.\begin{array}{l} \begin{Bmatrix}\hat{\boldsymbol{q}}_1(2)\\ \hat{\boldsymbol{q}}_2(2)\end{Bmatrix} = \begin{Bmatrix}(\hat{\sigma}_x,\ \hat{\sigma}_y,\ \hat{\sigma}_z)\\ (\hat{\sigma}_x,\ -\hat{\sigma}_z,\ \hat{\sigma}_y)\end{Bmatrix} \\ \xrightarrow{\textbf{cnot}} \begin{Bmatrix}\hat{\boldsymbol{q}}_1(2)\\ \hat{\boldsymbol{q}}_2(2)\end{Bmatrix} = \begin{Bmatrix}(\hat{\sigma}_x,\ \hat{\sigma}_z,\ -\hat{\sigma}_y)\\ (\hat{\sigma}_x,\ -\hat{\sigma}_z,\ \hat{\sigma}_y)\end{Bmatrix} \\ \xrightarrow{\textbf{not},1} \begin{Bmatrix}\hat{\boldsymbol{q}}_1(2)\\ \hat{\boldsymbol{q}}_2(2)\end{Bmatrix} = \begin{Bmatrix}(\hat{\sigma}_x,\ -\hat{\sigma}_z,\ \hat{\sigma}_y)\\ (\hat{\sigma}_x,\ -\hat{\sigma}_z,\ \hat{\sigma}_y)\end{Bmatrix} \end{array}\right\}. \tag{7.7}$$

While for any time $0 \leq t \leq 2$, the reference qubit is in the state $\hat{\boldsymbol{q}}_3(t) = \left(\hat{\sigma}_x,\ \hat{\sigma}_y,\ \hat{\sigma}_z\right)$. These expressions for the descriptors, combined with the system's Heisenberg state, completely specify the states of the qubits while they interact on the closed timelike curve. Thus, unorthodox qubits allow for a consistent solution of the grandfather paradox while neither classical physics nor orthodox quantum theory (in the Heisenberg picture) does.[16]

What has happened to the paradox? The paradox is avoided because $\mathfrak{Q}_2$'s $z$-observable is non-sharp, as can be deduced from (7.7) and the Heisenberg state. So, after the unorthodox **cnot** gate is applied and $\mathfrak{Q}_1$ has measured $\hat{q}_{2z}(0)$ at time $t = 1$, $\mathfrak{Q}_1$ is in a superposition of having received the messages 'it was a 1' and 'it was a $-1$'. In classical physics, a paradox arises because the system cannot be in both of those states simultaneously, but quantum theory does allow this, which is how the paradox is resolved.

[16] Notably, this resolution to the grandfather paradox is different from Deutsch's (*op. cit.*) treatment of the paradox. For instance, in his construction, the qubit leaves the closed timelike curve in a mixed state, whereas in my construction the qubit leaves it in a pure state.

This resolution corresponds to a situation in which an agent, Alice, travels back in time if she receives from her older self the message 'I did not travel back in time' whereas she does nothing if she receives 'I travelled back in time'. When the younger Alice receives this transmission from the older Alice, the younger Alice branches into two instances, each of which sees one of those messages. One of these instances of Alice will travel back in time and the other does not. Moreover, the version of Alice who does travel back in time will transmit a message that causes a younger Alice to do nothing; whereas the version of Alice who does not travel back in time transmits a message that causes a younger Alice to travel back in time. Hence, Alice's history while on the closed timelike curve is completely consistent.

### 7.2 Creating and destroying Hilbert space

Consider two qubits $\mathfrak{Q}_1$ and $\mathfrak{Q}_2$ on a closed timelike curve where, at a time $t = 1$, they experience a negative time delay of $-1$ units of time; the negative time-delay produces the older qubits $\mathfrak{Q}_3$ and $\mathfrak{Q}_4$ at the earlier time $t = 0$, so the qubits should adhere to the kinematic-consistency constraints

$$\left.\begin{aligned} \hat{\boldsymbol{q}}_3(0) &= \hat{\boldsymbol{q}}_1(1) \\ \hat{\boldsymbol{q}}_4(0) &= \hat{\boldsymbol{q}}_2(1) \end{aligned}\right\}. \tag{7.8}$$

Starting at this earlier time $t = 0$, the qubits $\mathfrak{Q}_1$, $\mathfrak{Q}_2$, $\mathfrak{Q}_3$ and $\mathfrak{Q}_4$ interact (figure 3). And due to these interactions, the Hilbert space of $\mathfrak{Q}_1$ and $\mathfrak{Q}_2$ does not necessarily have the same dimension as the Hilbert space of $\mathfrak{Q}_3$ and $\mathfrak{Q}_4$. For example, consider the case in which $\mathfrak{Q}_1$ and $\mathfrak{Q}_2$ maximally non-commute, so the Hilbert space of those qubits is 2-dimensional. Then, let the Hamiltonians of $\mathfrak{Q}_1$ and $\mathfrak{Q}_2$ between $t = 0$ and $t = 1$ be

$$\left.\begin{aligned} \hat{H}_1\big(\hat{\boldsymbol{q}}_1(t), \hat{\boldsymbol{q}}_2(t), \hat{\boldsymbol{q}}_3(t), \hat{\boldsymbol{q}}_4(t)\big) &= \frac{\pi}{4}\big\{\hat{q}_{1x}(t), \vec{P}\big(\hat{q}_{3z}(t), \hat{q}_{4z}(t)\big)\big\} \\ \hat{H}_2\big(\hat{\boldsymbol{q}}_1(t), \hat{\boldsymbol{q}}_2(t), \hat{\boldsymbol{q}}_3(t), \hat{\boldsymbol{q}}_4(t)\big) &= \frac{\pi}{4}\big\{\hat{q}_{2x}(t), \vec{P}\big(\hat{q}_{3z}(t), \hat{q}_{4z}(t)\big)\big\} \end{aligned}\right\}, \tag{7.9}$$

while the Hamiltonians of $\mathfrak{Q}_3$ and $\mathfrak{Q}_4$ are zero. Notably, all these Hamiltonians are time-independent between $t = 0$ and $t = 1$ because the terms $\hat{q}_{1x}(t)$ and $\hat{q}_{2x}(t)$ commute with their Hamiltonians and the descriptors of the other qubits are stationary.

An initial matrix representation for which the qubits satisfy the kinematic-consistency condition (7.8) when gate (7.9) is enacted, is as follows:

$$\begin{Bmatrix} \hat{\boldsymbol{q}}_1(0) \\ \hat{\boldsymbol{q}}_2(0) \\ \hat{\boldsymbol{q}}_3(0) \\ \hat{\boldsymbol{q}}_4(0) \end{Bmatrix} = \begin{Bmatrix} \big(\hat{\sigma}_x \otimes \hat{1}_2, & \hat{\sigma}_y \otimes \hat{1}_2, & \hat{\sigma}_z \otimes \hat{1}_2\big) \\ \big(\hat{\sigma}_x \otimes \hat{1}_2, & \hat{\sigma}_y \otimes \hat{1}_2, & \hat{\sigma}_z \otimes \hat{1}_2\big) \\ \big(\hat{\sigma}_x \otimes \hat{1}_2, & \hat{\sigma}_y \otimes \hat{\sigma}_z, & \hat{\sigma}_z \otimes \hat{\sigma}_z\big) \\ \big(\hat{\sigma}_x \otimes \hat{\sigma}_x, & \hat{\sigma}_y \otimes \hat{\sigma}_x & \hat{\sigma}_z \otimes \hat{1}_2\big) \end{Bmatrix}. \tag{7.10}$$

From this representation and the stationarity of $\mathfrak{Q}_3$ and $\mathfrak{Q}_4$ it follows that the Hamiltonians in (7.9) reduces to $\hat{H}_1 = \hat{H}_2 = \frac{\pi}{4}\hat{\sigma}_x \otimes (\hat{1}_2 - \hat{\sigma}_z)$, which produce the output states

$$\begin{Bmatrix} \hat{\boldsymbol{q}}_1(1) \\ \hat{\boldsymbol{q}}_2(1) \\ \hat{\boldsymbol{q}}_3(1) \\ \hat{\boldsymbol{q}}_4(1) \end{Bmatrix} = \begin{Bmatrix} (\hat{\sigma}_x \otimes \hat{1}_2, & \hat{\sigma}_y \otimes \hat{\sigma}_z, & \hat{\sigma}_z \otimes \hat{\sigma}_z) \\ (\hat{\sigma}_x \otimes \hat{\sigma}_x, & \hat{\sigma}_y \otimes \hat{\sigma}_x, & \hat{\sigma}_z \otimes \hat{1}_2) \\ (\hat{\sigma}_x \otimes \hat{1}_2, & \hat{\sigma}_y \otimes \hat{\sigma}_z, & \hat{\sigma}_z \otimes \hat{\sigma}_z) \\ (\hat{\sigma}_x \otimes \hat{\sigma}_x, & \hat{\sigma}_y \otimes \hat{\sigma}_x, & \hat{\sigma}_z \otimes \hat{1}_2) \end{Bmatrix}. \tag{7.11}$$

It can be readily verified that the output states (7.11) and input states (7.10) satisfy the self-consistency condition (7.8). Moreover, when the qubits $\mathfrak{Q}_3$ and $\mathfrak{Q}_4$ are considered in isolation, the q-numbers expressible algebraically from their descriptors $\hat{\boldsymbol{q}}_3(1)$ and $\hat{\boldsymbol{q}}_4(1)$ is spanned by the 16 basis-elements $\hat{1}_2 \otimes \hat{1}_2$, $\hat{\sigma}_i \otimes \hat{1}_2$, $\hat{1}_2 \otimes \hat{\sigma}_j$, and $\hat{\sigma}_k \otimes \hat{\sigma}_l$, where $i, j, k$, and $l$ range over the values in $\{x, y, z\}$. Consequently, the Hilbert space of $\mathfrak{Q}_3$ and $\mathfrak{Q}_4$ is 4-dimensional, whereas the Hilbert space of $\mathfrak{Q}_1$ and $\mathfrak{Q}_2$ is only 2-dimensional, so the Hilbert space in the unambiguous future of the qubits is larger than it is in their unambiguous past.[17]

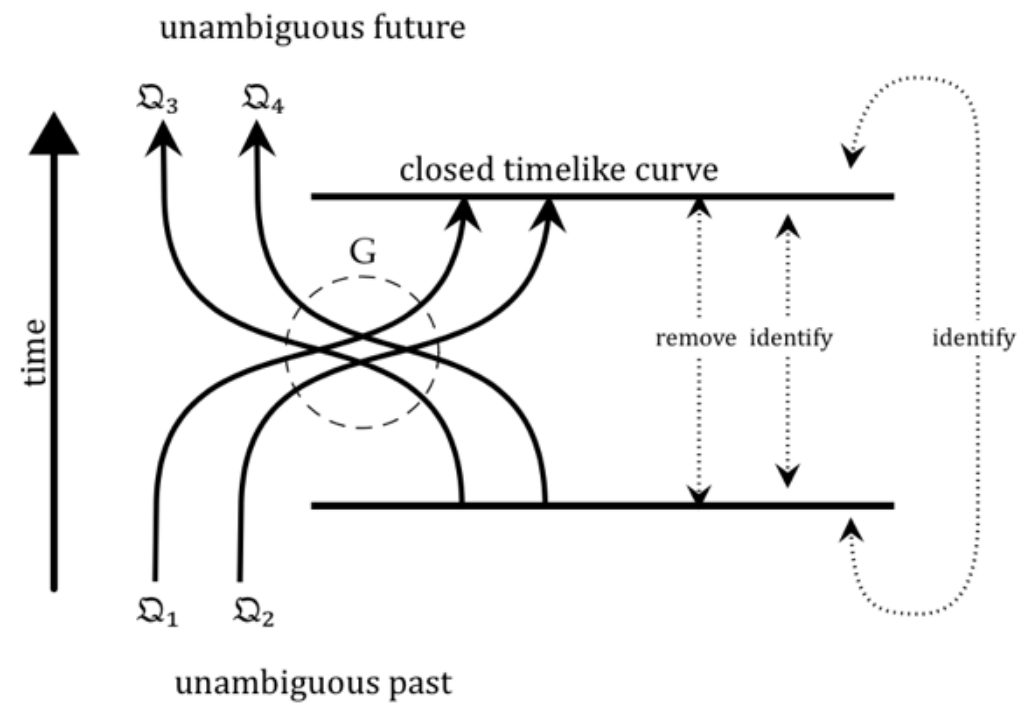

Figure 3. A spacetime diagram depicting the world lines of $\mathfrak{Q}_1$,$\mathfrak{Q}_2$,$\mathfrak{Q}_3$ and $\mathfrak{Q}_4$ and interaction region G, as well as the qubits' unambiguous past and future.

The above-described protocol can also be applied in reverse. In that case, the qubits' Hilbert space is smaller in the unambiguous future than it is in the unambiguous past so that the state space of the qubits is comparatively smaller in the unambiguous future (see §2.3). Hence, if such Hilbert-space destroying processes occur in, for instance, the interior of a black hole, then perhaps this protocol, and protocols like it, provide a mechanism through which black holes can erase information stored in the state of the initial qubits $\mathfrak{Q}_1$ and $\mathfrak{Q}_2$. I will leave this question open for future research.

[17] In §2.2, I demonstrated that unorthodox qubits have a Hilbert space of constant dimension. However, it is only isolated systems that have an invariant Hilbert space. The two-qubit subsystem consisting of $\mathfrak{Q}_1$ and $\mathfrak{Q}_2$ is not isolated and so its Hilbert space is not necessarily a constant of the motion.

## 8 Conclusions

I have elaborated some implications of a locally realistic, 'unorthodox' variant of quantum theory (following Deutsch (*op. cit.*)) in which the descriptors of spatially separate systems do not necessarily commute. I have shown that unorthodox qubits support novel types of gate, that they can instantiate classical information, that they allow for a resolution of the grandfather paradox in the presence of closed timelike curves, and that they can create and destroy Hilbert space.

### Acknowledgements

I am grateful for the many discussions with David Deutsch, Chiara Marletto, and Vlatko Vedral, who helped me understand the material in this paper better than I could have done alone. I also wish to thank David Deutsch, Chiara Marletto, Eric Marcus, and Abel Jansma for their useful comments on earlier manuscripts of this paper. This work was supported in part by the Prins Bernhard Cultuurfonds.